\documentclass[12pt]{article}

\usepackage{multirow}

\usepackage[letterpaper,left=1.25in,right=1.25in,top=1in,bottom=1in]{geometry}

\usepackage[T1]{fontenc}
\usepackage[utf8]{inputenc}
\usepackage{newtxtext,newtxmath} 

\usepackage{setspace}
\usepackage{titlesec}
\titleformat{\section}{\bfseries\fontsize{16}{18}\selectfont}{\thesection}{1em}{}
\titlespacing*{\section}{0pt}{2\baselineskip}{1\baselineskip}

\titleformat{\subsection}{\bfseries\fontsize{14}{16}\selectfont}{\thesubsection}{1em}{}
\titlespacing*{\subsection}{0pt}{2\baselineskip}{1\baselineskip}

\titleformat{\subsubsection}{\bfseries\fontsize{12}{14}\selectfont}{\thesubsubsection}{1em}{}
\titlespacing*{\subsubsection}{0pt}{2\baselineskip}{1\baselineskip}

\usepackage{fancyhdr}
\usepackage{graphicx}
\usepackage{float}
\usepackage{booktabs}
\usepackage{caption}
\usepackage{subcaption}
\usepackage[numbers]{natbib}

\usepackage[colorlinks=true,linkcolor=black,citecolor=black,urlcolor=black]{hyperref}

\newcommand{\MICSTitle}[1]{%
  \begin{center}
    \vspace*{1.5in} 
    {\fontsize{18}{20}\selectfont #1\par}
    \vspace{2\baselineskip} 
  \end{center}
}

\newcommand{\MICSAuthorBlock}[5]{%
  \begin{center}
    {\fontsize{14}{16}\selectfont #1\par} 
    {\fontsize{14}{16}\selectfont #2\par} 
    {\fontsize{14}{16}\selectfont #3\par} 
    {\fontsize{14}{16}\selectfont #4\par} 
    {\fontsize{14}{16}\selectfont #5\par} 
    \vspace{2\baselineskip} 
  \end{center}
}

\newcommand{\MICSAbstract}[1]{%
  \begin{center}
    {\bfseries\fontsize{16}{18}\selectfont Abstract\par}
  \end{center}
  \vspace{\baselineskip} 
  {\fontsize{12}{14}\selectfont
  #1\par}
}

\begin{document}

\thispagestyle{empty}

\MICSTitle{Multimodal Image Colorization: Quantifying the Impact of Text-Conditioned Guidance on Grayscale-to-Color Translation}

\MICSAuthorBlock
  {Colten Reissmann, Hugo Garrido-Lestache Belinchon}
  {Department of Computer Science and Software Engineering}
  {Milwaukee School of Engineering}
  {Milwaukee, WI US}
  {\{reissmannc, garrido-lestacheh\}@msoe.edu}

\MICSAbstract{
Grayscale images are commonly found in historical photography restoration, medical imaging, and artistic media. However, automatically applying color to these images has proven to be a significant challenge in computer vision, as there are many valid ways to color an image.

In this work, we quantify the effect of text conditioning on the pixel-level and perceptual metrics of grayscale-to-color image models. Specifically, we compare two architectures (a U-Net and Stable Diffusion 1.5), each tested with and without CLIP text conditioning while holding everything else equal. Our results show that text conditioning improves PSNR by 5.6\%, SSIM by 1.2\%, and colorfulness by 36.6\%, while reducing LPIPS by 7.6\% in the U-Net tier, and improves PSNR by 5.8\%, SSIM by 1.5\%, and colorfulness by 0.6\%, while reducing LPIPS by 11.3\% in the Stable Diffusion tier. We conclude that text conditioning provides consistent, measurable improvement to colorization quality across both architecture scales.
}

\setcounter{page}{1}
\pagestyle{fancy}

\section{Introduction}
\subsection{The Colorization Problem}
Grayscale images are common in a variety of domains including historical photography restoration, medical imaging, and artistic media. Automatic image colorization aims to transform grayscale input images into visually plausible color images while preserving structure and semantics. However, this task is fundamentally ill-posed because grayscale intensity does not uniquely determine the original color values. In many cases, a single grayscale image may correspond to many different, visually acceptable colorizations depending on the context of the scene, the identity of the object, or artistic interpretation. As a result, successful colorization systems must infer semantic information about objects and environments in the image to produce realistic colors instead of simply mapping pixel intensities to fixed color values. Modern deep-learning approaches attempt to address this ambiguity using probabilistic models, semantic feature extraction, and generative architectures, but the inherent multimodality of the problem remains a central challenge in automatic colorization research \cite{10.3389/fcomp.2025.1626641}. 

\subsection{Text as Disambiguation}
One promising strategy to reduce color ambiguity is to incorporate natural-language descriptions as an additional conditioning signal during the colorization process. Text can explicitly define attributes that cannot be determined from grayscale intensity alone, for example, distinguishing between a ``red car'' and a ``blue car'', which would otherwise appear identical in grayscale. Several recent methods integrate textual prompts or captions to guide the colorization process. Systems such as TIC \cite{ghosh2022tictextguidedimagecolorization}, DiffColor \cite{lin2023diffcolorhighfidelitytextguided}, and Diffusing Colors \cite{zabari2023diffusingcolorsimagecolorization} demonstrate that incorporating textual conditioning can improve perceptual quality and semantic correctness by aligning generated colors with language descriptions. However, these approaches introduce multiple simultaneous architectural changes, such as modified backbones, diffusion-based generative frameworks, and new loss formulations, alongside the addition of text conditioning. As a result, improvements reported in these works cannot be attributed solely to the inclusion of text guidance, making it difficult to isolate the specific contribution of natural-language information to colorization performance.

\subsection{Research Question and Contributions}
The quantitative benefits of text conditioning in the context of grayscale image colorization have been under-investigated. This paper investigates the research question: \emph{To what extent does text conditioning improve automatic image colorization quality compared to an identical model without text input?} To answer this, we make three contributions: (1) a controlled ablation study on a U-Net architecture and a Stable Diffusion 1.5 architecture where the only variable is the presence of text conditioning; (2) quantitative evaluation using PSNR, SSIM, LPIPS, and colorfulness on held-out test data; and (3) qualitative analysis of when text conditioning helps most and when it provides limited advantage.

\section{Related Work}
Early deep-learning colorization methods predict chrominance from grayscale input using CNNs. Zhang et al.~\citep{DBLP:journals/corr/ZhangIE16} frame colorization as classification in CIE-Lab space with class-rebalancing, while Iizuka et al.~\citep{IizukaSIGGRAPH2016} merge local and global features to predict ab channels. These methods optimize L1/L2 losses, sometimes augmented by perceptual \cite{DBLP:journals/corr/JohnsonAL16} or adversarial losses such as the pix2pix PatchGAN framework \citep{DBLP:journals/corr/IsolaZZE16}. GAN losses sharpen colors at the risk of artifacts, while L1/L2 losses produce blurrier, desaturated results.

Recent work extends colorization by introducing textual guidance. Ghosh et al. (TIC, 2022) introduce one of the first text-conditioned image colorizers, feeding a grayscale image along with a text prompt into a GAN. The caption is encoded alongside image features, providing color cues (e.g., ``red apple'') to improve fidelity. In experiments, TIC achieved better SSIM, PSNR, and LPIPS scores than prior unconditional methods, validating the idea that text hints improve color accuracy \citep{ghosh2022tictextguidedimagecolorization}. Lin et al. (2024) propose DiffColor, which fine-tunes a pretrained text-to-image diffusion model with a CLIP-based contrastive loss to align the colorized output with the prompt. They demonstrate that simple prompt changes can produce diverse yet semantically consistent recolorings (e.g., ``red car'' vs.~``blue car''), and their user study confirms that DiffColor outperforms previous methods in visual quality and diversity \citep{lin2023diffcolorhighfidelitytextguided}.
Importantly, all of these text-guided methods introduce architectural changes alongside text conditioning, including new backbones, loss functions, or generative frameworks, which makes it impossible to attribute their improvements to text alone. Our work addresses this gap by holding all other variables constant.

Diffusion models have recently been applied to colorization with strong results. Saharia et al. (2022) develop \textit{Palette}, a conditional diffusion framework for image-to-image tasks like colorization. Palette trains on multiple translation tasks without task-specific losses and outperforms strong GAN and regression baselines \citep{saharia2022paletteimagetoimagediffusionmodels}. Rombach et al. (2022) introduce latent diffusion models, which operate in a compressed latent space for efficiency; the publicly released Stable Diffusion 1.5 checkpoint serves as the backbone for our SD-tier experiments \cite{rombach2022high}. Zabari et al. (2023) present ``Diffusing Colors,'' which fine-tunes Stable Diffusion on auto-generated captions while incorporating a CLIP-based ranker to control colorfulness \citep{zabari2023diffusingcolorsimagecolorization}. Hang et al. (2023) introduce Min-SNR-$\gamma$ weighting to balance diffusion training across timesteps, achieving approximately 3.4$\times$ faster convergence \citep{hang2024efficientdiffusiontrainingminsnr}.

Radford et al.~\cite{radford2021learning} introduce CLIP, which uses contrastive pretraining on image--text pairs to produce aligned visual and textual representations widely adopted as the conditioning signal in text-guided generation. Cross-attention, derived from the transformer architecture \cite{attentionneed}, enables image features (queries) to selectively attend to text embeddings (keys/values). Ho and Salimans \cite{ho2022classifier} propose classifier-free guidance (CFG), which extrapolates between unconditional and conditional predictions at inference using a guidance scale variable, where a higher guidance scale amplifies text but risks artifacts. Our SD-tier experiments use a guidance scale of 1.0 to isolate text's effect through cross-attention alone.

Many colorization systems operate in CIELAB space, predicting only the chrominance (ab) channels while preserving the luminance (L), which simplifies the task by fixing brightness and preserving image structure \citep{DBLP:journals/corr/ZhangIE16, IizukaSIGGRAPH2016}. Evaluation typically combines pixel-level metrics (PSNR, SSIM \cite{wang2004image}), perceptual distance (LPIPS \cite{zhang2018unreasonable}), and color vividness (Hasler--S\"usstrunk colorfulness \cite{hasler2003measuring}).

\section{Methodology}

\subsection{Dataset}

All experiments use the \textit{Open Image Preferences v1} dataset \cite{openimageprefs} from HuggingFace (data-is-better-together/open-image-preferences-v1-binarized): 7{,}459 prompt--image preference pairs spanning diverse visual categories. Each prompt describes the desired image content and often includes color-relevant attributes (e.g., ``vibrant colors,'' ``blue knight,'' ``glacier landscape''). The raw Parquet files are preprocessed into aligned grayscale/color PNG pairs and a JSON caption index mapping filenames to prompts.

A deterministic train/test split allocates 90\% of images for training (6{,}714) and 10\% for evaluation (745), with a further 90/10 train/validation split within the training set. Both the U-Net and Stable Diffusion tiers use the same split, ensuring no test-set leakage in either tier.

\begin{figure}[H]
  \centering
  \includegraphics[width=\linewidth]{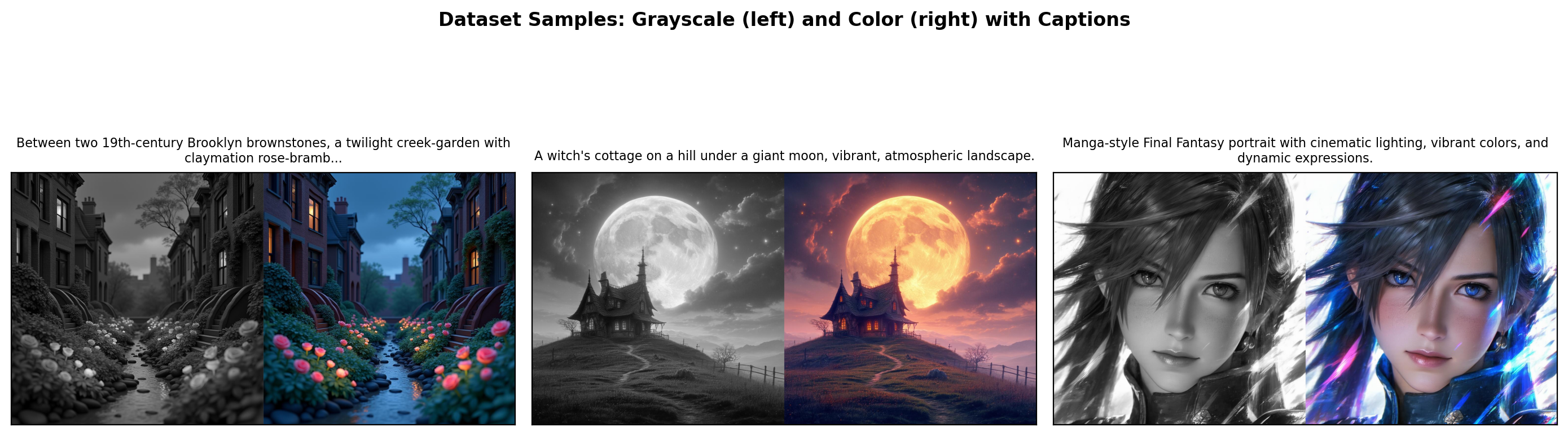}
  \caption{Dataset samples showing grayscale (left) and color (right) pairs with their text captions. Captions contain color-relevant descriptions that guide the colorization process.}
  \label{fig:dataset_samples}
\end{figure}

\subsection{Architecture Tier 1 --- U-Net}

Both U-Net variants share a custom backbone that maps a single-channel grayscale image to a three-channel RGB image at $256 \times 256$. The encoder has five stages ($64 \to 128 \to 256 \to 512 \to 512$ channels), each consisting of two convolution--batch-normalization--ReLU blocks followed by $2{\times}2$ max pooling, with graduated dropout. The decoder mirrors this with four bilinear-upsampling stages and skip connections. A $1{\times}1$ convolution and sigmoid activation produce the output. Both variants use MSE loss, Adam ($\text{lr}=10^{-4}$), and learning-rate scheduling; full hyperparameters are listed in Table~\ref{tab:hyperparams}.

\textbf{UN-NP (unconditioned)} receives only the grayscale image ($\sim$13.4M parameters, batch size 16).

\textbf{UN-P (text-conditioned)} adds cross-attention layers at encoder depths 3--5 and decoder depths 1--2, where image features attend to embeddings from a frozen CLIP ViT-B/32 encoder \cite{radford2021learning} via 8-head multi-head attention with residual connections. Attention is omitted at shallow layers where features are high-resolution and low-semantic. During training, 10\% of captions are replaced with empty strings. The batch size is reduced to 8 for memory; the model has $\sim$16.4M parameters ($\sim$3.1M from attention layers). All other settings match UN-NP.

\subsection{Architecture Tier 2 --- Stable Diffusion 1.5}

Both SD variants fine-tune the Stable Diffusion 1.5 checkpoint \cite{rombach2022high} with identical hyperparameters. The pipeline comprises a frozen VAE encoder (compressing $512{\times}512$ images to $64{\times}64$ latents), a frozen CLIP ViT-L/14 text encoder (768-dim), and a trainable denoising UNet with $\sim$859M parameters. The UNet's input convolution is expanded from 4 to 8 channels to concatenate grayscale latents alongside noisy color latents; pretrained weights are preserved for the original channels and new channels are zero-initialized.

The model operates in CIELAB space, predicting ab chrominance while preserving the input luminance. Training uses $\epsilon$-prediction MSE loss with Min-SNR-$\gamma$ weighting ($\gamma = 5.0$) \cite{hang2024efficientdiffusiontrainingminsnr}, EMA (decay $= 0.9999$), noise offset of 0.1, mixed precision (fp16), gradient checkpointing, and a cosine schedule with 500-step warmup. Inference uses DDIM with 50 steps. Full hyperparameters are in Table~\ref{tab:hyperparams}.

Both variants use guidance scale $= 1.0$ (no CFG amplification) \cite{ho2022classifier}. Earlier experiments showed that high CFG scales caused the text signal to overwhelm the grayscale conditioning, producing overly vivid or incorrect colors. Setting the guidance scale to 1.0 isolates text's effect through cross-attention alone.

\textbf{SD-NP (no text)} was trained with every caption replaced by an empty string, so the model never receives meaningful text.

\textbf{SD-P (text-conditioned)} was trained with every sample using its real caption. All other settings are identical to SD-NP.

\subsection{Evaluation Metrics}

We evaluate with four metrics computed between predicted and ground-truth images:
\textbf{PSNR} (peak signal-to-noise ratio; higher is better) and \textbf{SSIM} (structural similarity; higher is better) \cite{wang2004image} measure pixel-level and structural fidelity.
\textbf{LPIPS} (learned perceptual similarity with AlexNet; lower is better) \cite{zhang2018unreasonable} captures perceptual quality.
\textbf{Colorfulness} (Hasler--S\"usstrunk index; higher is more vivid) \cite{hasler2003measuring} measures saturation.

\subsection{Experimental Setup}

All training and evaluation is conducted on an NVIDIA DGX-1 GPU cluster via SLURM. Metrics are computed on the 745-image held-out test set. To enable fair cross-architecture comparison, all metrics are computed at $512 \times 512$: SD models run at this resolution by default, while U-Net predictions ($256^2$) are upscaled using bilinear interpolation before metric computation. This penalizes U-Net scores slightly but ensures consistent evaluation. Table~\ref{tab:hyperparams} summarizes the configuration of each variant.

\begin{table}[H]
\centering
\caption{Training configuration for each model variant.}
\resizebox{\linewidth}{!}{%
\begin{tabular}{lcccc}
\toprule
                           & \textbf{UN-NP} & \textbf{UN-P} & \textbf{SD-NP} & \textbf{SD-P} \\
\midrule
Architecture               & U-Net        & U-Net        & SD 1.5           & SD 1.5          \\
Parameters                 & $\sim$13.4M  & $\sim$16.4M  & $\sim$859M       & $\sim$859M      \\
Color space                & RGB          & RGB          & LAB (ab)         & LAB (ab)        \\
Resolution                 & $256^2$      & $256^2$      & $512^2$          & $512^2$         \\
Text encoder               & ---          & CLIP ViT-B/32 & --- (empty)     & CLIP ViT-L/14   \\
Loss                       & MSE          & MSE          & $\epsilon$-MSE + Min-SNR & $\epsilon$-MSE + Min-SNR \\
Optimizer                  & Adam         & Adam         & AdamW            & AdamW           \\
Learning rate              & $10^{-4}$    & $10^{-4}$    & $5{\times}10^{-5}$ & $5{\times}10^{-5}$ \\
Batch size                 & 16           & 8            & 4                & 4               \\
Epochs trained             & 60           & 95           & 50               & 50              \\
Guidance scale             & ---          & ---          & 1.0              & 1.0             \\
Empty prompt prob.         & ---          & 10\%         & 100\%            & 0\%             \\
\bottomrule
\end{tabular}%
}
\label{tab:hyperparams}
\end{table}

\section{Results}

\subsection{U-Net: UN-NP vs UN-P}

Table~\ref{tab:cross_tier} (Section~\ref{sec:cross_tier}) presents the full cross-tier comparison.
UN-P improved over UN-NP by +5.6\% PSNR (19.13$\to$20.21), +1.2\% SSIM, $-$7.6\% LPIPS (0.376$\to$0.348), and +36.6\% colorfulness.

During the training process, UN-NP reached its best validation MSE of 0.01197 at epoch 41 of 60 and could not improve further despite continued training. UN-P consistently achieved lower validation loss, reaching a best MSE of 0.01038 at epoch 76, which is a 13.3\% reduction over UN-NP's best.

\begin{figure}[H]
 \centering
 \begin{subfigure}[b]{0.45\linewidth}
   \includegraphics[width=\linewidth]{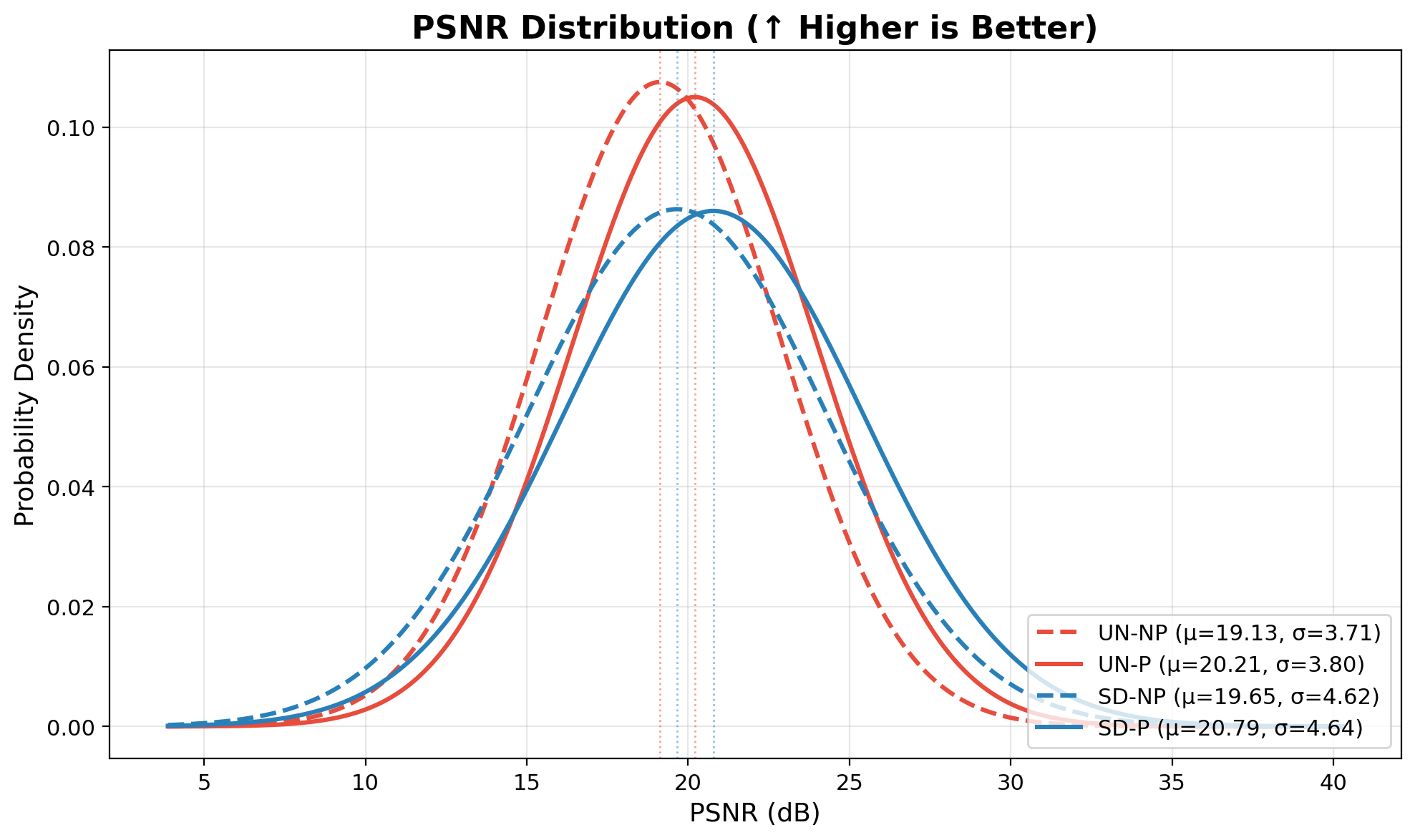}
   \caption{PSNR}
 \end{subfigure}
 \hfill
 \begin{subfigure}[b]{0.45\linewidth}
   \includegraphics[width=\linewidth]{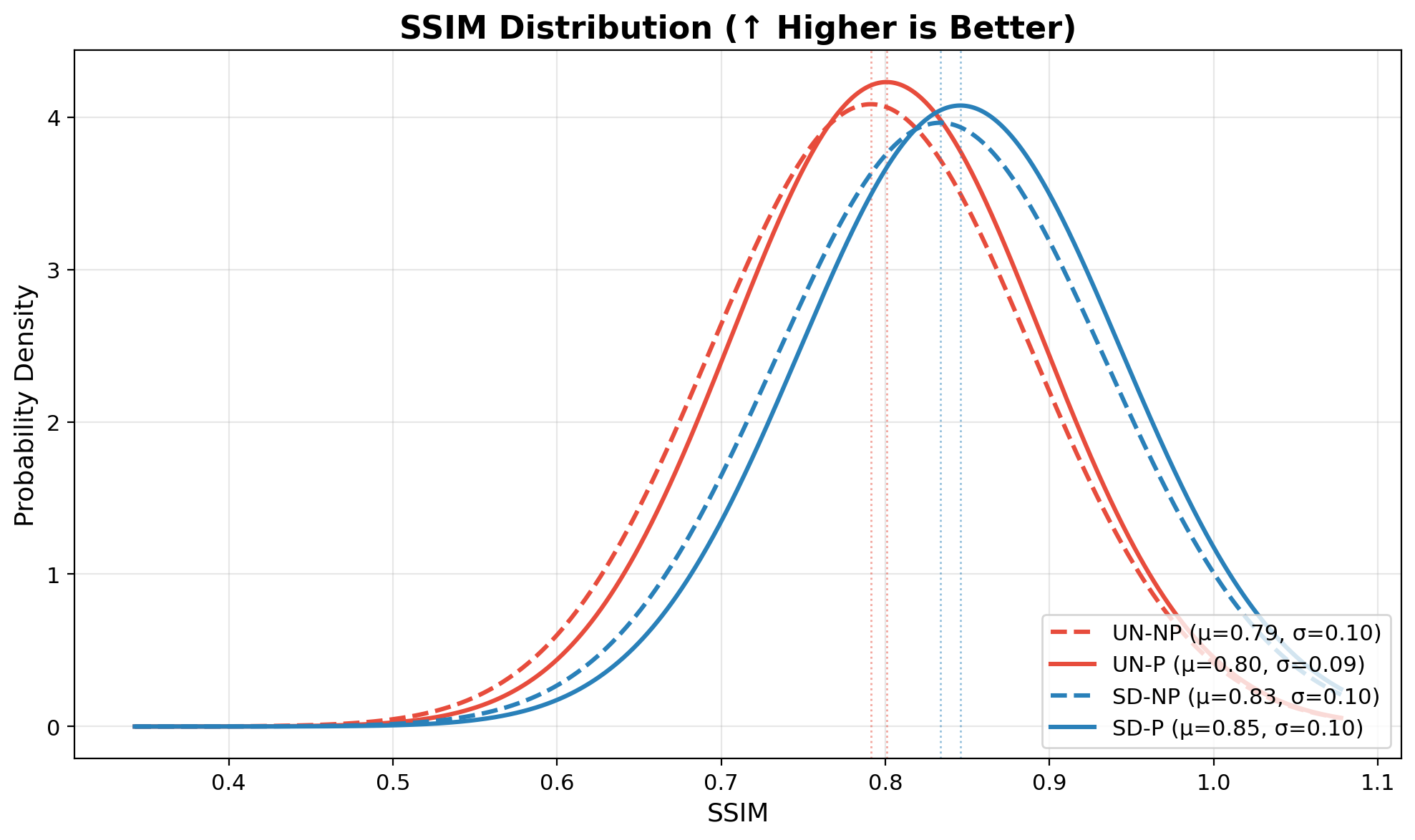}
   \caption{SSIM}
 \end{subfigure}
 \hfill
 \newline
 \begin{subfigure}[b]{0.45\linewidth}
   \includegraphics[width=\linewidth]{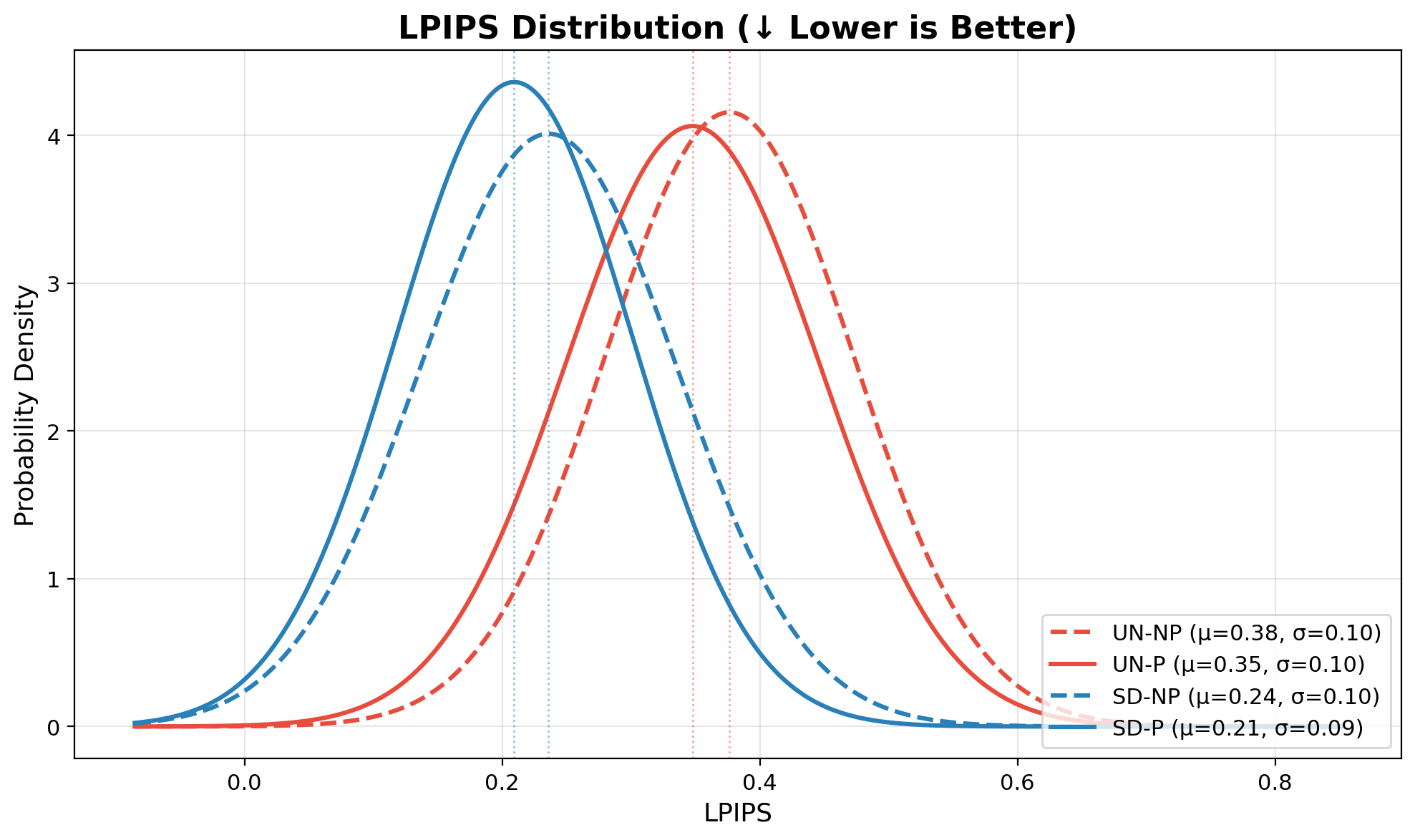}
   \caption{LPIPS}
 \end{subfigure}
 \caption{Per-image metric distributions for all four models. Text conditioning shifts PSNR and SSIM right (better) and LPIPS left (better).}
 \label{fig:metric_distributions}
\end{figure}

\begin{figure}[H]
  \centering
  \includegraphics[width=0.85\linewidth]{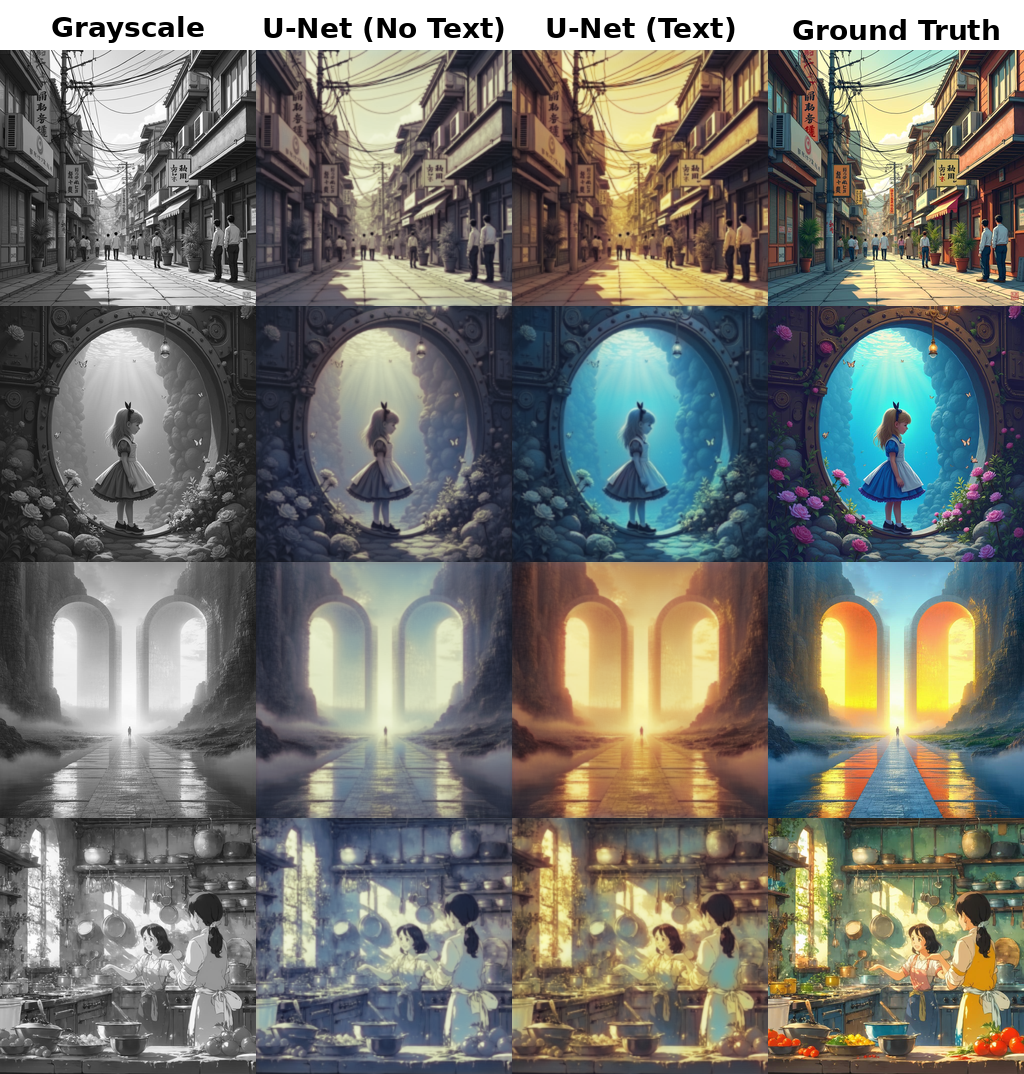}
  \caption{Qualitative U-Net results. From left to right: grayscale input, UN-NP (unconditioned), UN-P (text-conditioned), ground truth.}
  \label{fig:unet_qualitative}
\end{figure}

\subsection{Stable Diffusion: SD-NP vs SD-P}

Table~\ref{tab:cross_tier} (Section~\ref{sec:cross_tier}) also shows the cross-tier comparison of the SD models.
SD-P improved over SD-NP by +5.8\% PSNR (19.65$\to$20.79), +1.5\% SSIM, $-$11.3\% LPIPS (0.236$\to$0.209), and +0.6\% colorfulness.

During the training process, SD-P achieved its best validation performance at epoch 31 (PSNR = 20.30, SSIM = 0.8624, LPIPS = 0.1872), while SD-NP later peaked at epoch 45 (PSNR = 18.92, SSIM = 0.8456, LPIPS = 0.2147). Comparing these best validation performances, there were relative improvements of +7.3\% PSNR, +2.0\% SSIM, and $-$12.8\% LPIPS when adding text conditioning.


\begin{figure}[H]
  \centering
  \includegraphics[width=\linewidth]{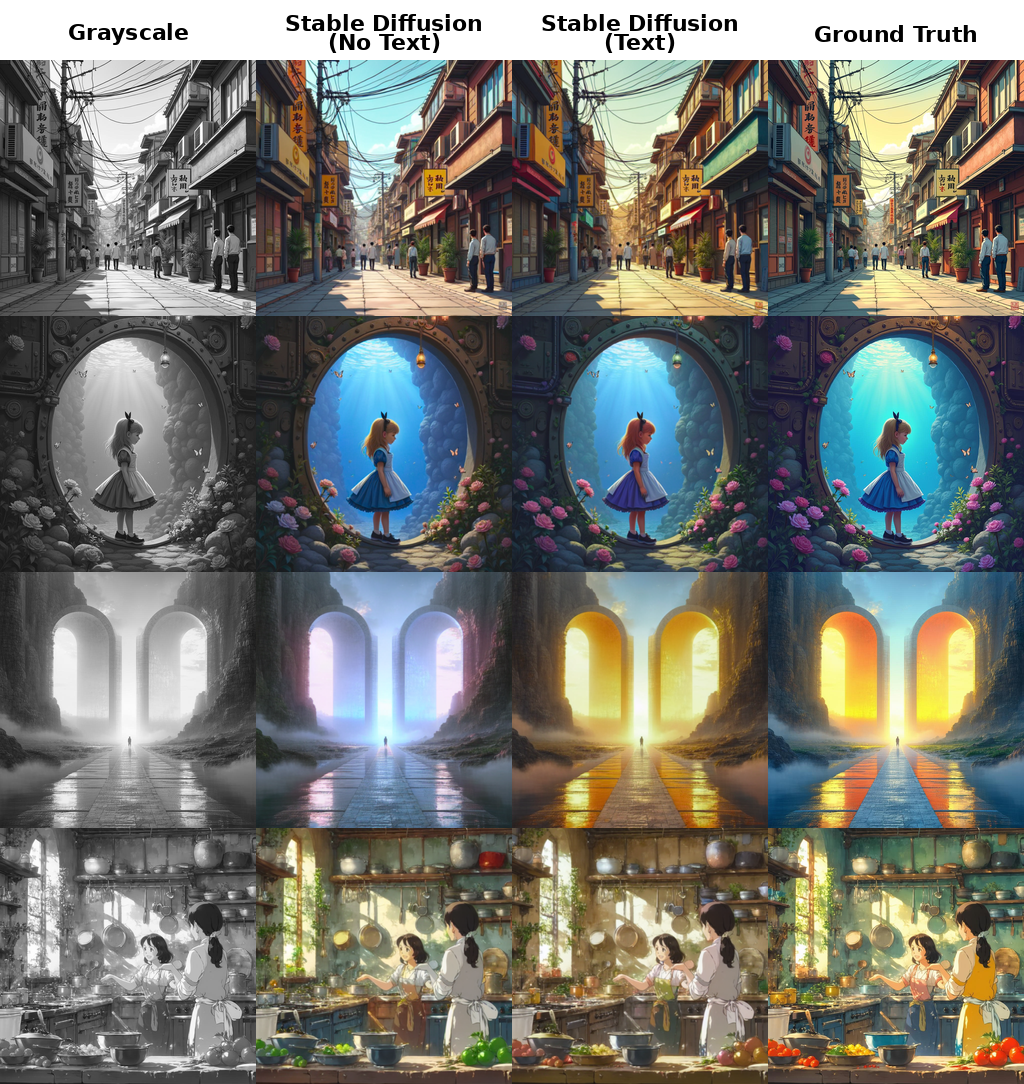}
  \caption{Qualitative Stable Diffusion results. From left to right: grayscale input, SD-NP (no text), SD-P (text-conditioned), ground truth.}
  \label{fig:sd_qualitative}
\end{figure}

\subsection{Cross-Tier Summary}
\label{sec:cross_tier}

\begin{table}[H]
\centering
\caption{Cross-tier summary. $\Delta$ rows show the relative change from adding text within each tier. All metrics computed at $512^2$; U-Net predictions are upscaled from $256^2$ via bilinear interpolation. Target colorfulness is 0.258.}
\begin{tabular}{llccccc}
\toprule
Tier & Model & Params & PSNR $\uparrow$ & SSIM $\uparrow$ & LPIPS $\downarrow$ & Colorfulness $\uparrow$ \\
\midrule
\multirow{3}{*}{U-Net}
  & UN-NP (no text)    & $\sim$13.4M & 19.13 & 0.7912 & 0.3763 & 0.093 \\
  & UN-P (text)       & $\sim$16.4M & 20.21 & 0.8008 & 0.3477 & 0.127 \\
  & $\Delta$ U-Net   & +22.8\%     & +5.6\% & +1.2\% & $-$7.6\% & +36.6\% \\
\midrule
\multirow{3}{*}{SD 1.5}
  & SD-NP (no text) & $\sim$859M  & 19.65 & 0.8334 & 0.2360 & 0.167 \\
  & SD-P (text)     & $\sim$859M  & 20.79 & 0.8457 & 0.2093 & 0.168 \\
  & $\Delta$ SD      & 0\%         & +5.8\% & +1.5\% & $-$11.3\% & +0.6\% \\
\bottomrule
\end{tabular}
\label{tab:cross_tier}
\end{table}

Text conditioning improved all four metrics in both tiers (Table~\ref{tab:cross_tier}). The PSNR improvement is comparable across tiers (+5.6\% U-Net vs.\ +5.8\% SD), but the SD tier achieves a larger LPIPS reduction ($-$11.3\% vs.\ $-$7.6\%). The U-Net tier shows a striking +36.6\% colorfulness increase, which suggests that text conditioning helps the from-scratch model produce substantially more vivid colors.


\subsection{Statistical Significance}
\label{sec:stat_sig}

Because both models in each tier are evaluated on the same 745 test images, the per-image metrics form naturally paired observations: for each image $i$, we observe both the NP score $x_i$ and the P score $y_i$, and compute the difference $d_i = y_i - x_i$. We test significance using two-sided paired $t$-tests and report Cohen's $d$ as the effect size. The paired $t$-test is the most powerful test for this design for two reasons. First, it exploits the paired structure: an independent (unpaired) $t$-test would treat the NP and P scores as separate samples, ignoring the per-image correspondence and inflating error variance with between-image variability (e.g., near-monochrome scenes are inherently easier than multi-object scenes). By modeling the within-image difference $d_i$ directly, the paired test removes this nuisance variance and yields tighter confidence intervals. Second, under the assumption that the paired differences are approximately normally distributed---an assumption supported by the per-image metric distributions in Figure~\ref{fig:metric_distributions}---the paired $t$-test is the uniformly most powerful unbiased test for detecting a mean shift; no non-parametric alternative (e.g., Wilcoxon signed-rank) can achieve higher power when normality holds. Table~\ref{tab:significance} summarizes the results.

\begin{table}[H]
\centering
\caption{Paired $t$-tests for NP $\to$ P within each tier ($n = 745$). All $p$-values are two-sided. $^{\ast\ast\ast}$$p < 0.001$;\; n.s.\ = not significant.}
\resizebox{\linewidth}{!}{%
\begin{tabular}{llrrrrl}
\toprule
Tier & Metric & Mean $\Delta$ & 95\% CI & Cohen's $d$ & $t$ & $p$ \\
\midrule
\multirow{4}{*}{U-Net}
  & PSNR $\uparrow$         & $+1.081$ & $[+0.944,\;+1.219]$ & $+0.57$ & $15.46$ & $5.5 \times 10^{-47}$$^{\ast\ast\ast}$ \\
  & SSIM $\uparrow$         & $+0.010$ & $[+0.008,\;+0.011]$ & $+0.40$ & $11.04$ & $2.3 \times 10^{-26}$$^{\ast\ast\ast}$ \\
  & LPIPS $\downarrow$      & $-0.029$ & $[-0.032,\;-0.025]$ & $-0.59$ & $-16.10$ & $2.8 \times 10^{-50}$$^{\ast\ast\ast}$ \\
  & Colorfulness $\uparrow$ & $+0.034$ & $[+0.030,\;+0.037]$ & $+0.67$ & $18.39$ & $1.5 \times 10^{-62}$$^{\ast\ast\ast}$ \\
\midrule
\multirow{4}{*}{SD 1.5}
  & PSNR $\uparrow$         & $+1.132$ & $[+0.927,\;+1.337]$ & $+0.40$ & $10.83$ & $1.7 \times 10^{-25}$$^{\ast\ast\ast}$ \\
  & SSIM $\uparrow$         & $+0.012$ & $[+0.009,\;+0.015]$ & $+0.28$ & $7.76$  & $2.7 \times 10^{-14}$$^{\ast\ast\ast}$ \\
  & LPIPS $\downarrow$      & $-0.027$ & $[-0.031,\;-0.022]$ & $-0.40$ & $-11.03$ & $2.7 \times 10^{-26}$$^{\ast\ast\ast}$ \\
  & Colorfulness $\uparrow$ & $+0.001$ & $[-0.004,\;+0.006]$ & $+0.01$ & $0.38$  & $7.1 \times 10^{-1}$\;n.s. \\
\bottomrule
\end{tabular}%
}
\label{tab:significance}
\end{table}

All improvements from text conditioning are highly significant ($p < 10^{-13}$) for PSNR, SSIM, and LPIPS in both tiers, with medium effect sizes (Cohen's $d$ ranging from 0.28 to 0.59). The sole exception is SD colorfulness, where the $+0.6\%$ relative increase reported in Table~\ref{tab:cross_tier} is not statistically significant ($p = 0.71$, $d = 0.01$), confirming that the SD backbone already saturates its colorfulness with or without text. In the U-Net tier, colorfulness shows the largest effect size ($d = 0.67$), consistent with our hypothesis that text provides the greatest marginal benefit when the model lacks pretrained color priors.

\subsection{Qualitative Analysis}

Beyond aggregate metrics, individual examples reveal when text conditioning helps most and when it provides little benefit.

\begin{figure}[H]
  \centering
  \includegraphics[width=\linewidth]{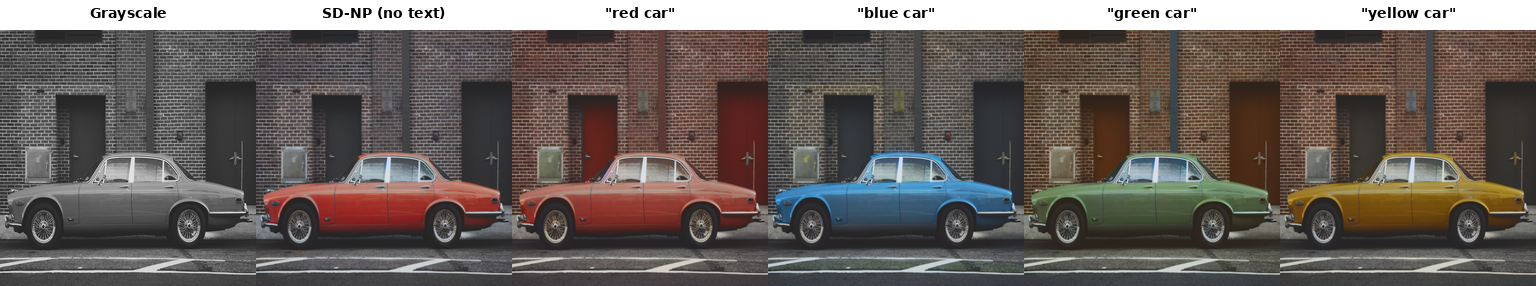}
  \caption{The same grayscale car colorized by SD-NP (no text) and by SD-P with four different color prompts. Text conditioning steers the model toward the specified color, demonstrating disambiguation of an otherwise ambiguous input.}
  \label{fig:car_prompt}
\end{figure}

\textbf{Text resolves ambiguity.} For objects with many plausible colorings (such as vehicles, clothing, and food), the unconditioned models frequently default to desaturated or averaged colors, while the text-conditioned variants produce colors consistent with the caption. Figure~\ref{fig:car_prompt} demonstrates this with a vintage car: when SD-NP receives only the grayscale input, it must guess the car's color. When SD-P receives prompts specifying ``red,'' ``blue,'' ``green,'' or ``yellow,'' it reasonably renders each variant. This illustrates how text disambiguates cases where grayscale intensity alone leaves the color underdetermined.

\textbf{Canonical scenes gain less.} Landscapes with stereotypical coloring show smaller differences between conditioned and unconditioned outputs, as both models can infer dominant colors from structure alone (Figures~\ref{fig:unet_qualitative} and~\ref{fig:sd_qualitative}).

\textbf{Artifact differences.} The U-Net tier produces blurry, desaturated outputs (a consequence of MSE loss), while the SD tier produces sharper results with higher perceptual fidelity. Text conditioning does not eliminate the U-Net's blurriness, but it does improve color accuracy within each tier.

\section{Discussion}

As shown in Table~\ref{tab:cross_tier}, text conditioning consistently improved all four metrics in both tiers. The consistency of this finding across architectures that differ by roughly $50{\times}$ in parameter count ($\sim$13--16M vs.\ $\sim$859M) strengthens the claim that text conditioning provides architecture-independent benefit.

While both tiers show similar PSNR gains, the SD tier achieves a larger LPIPS reduction while the U-Net shows a dramatically larger colorfulness increase. We hypothesize that text conditioning provides a larger relative improvement for the U-Net because this architecture lacks pretrained visual priors. Trained from scratch on only 6{,}714 images, the U-Net must learn all color semantics entirely from this limited data, and text provides an external source of semantic information that the model cannot extract from grayscale pixels alone. In contrast, the Stable Diffusion backbone was pretrained on billions of text--image pairs, which already encodes rich associations between visual content and color, so explicit text conditioning provides less marginal information.

Across multiple iterations of the U-Net architecture (V1, V2, and the predecessor of UN-NP), validation MSE consistently plateaued between 0.0135 and 0.0155 regardless of changes to regularization, dropout rates, or a five-fold increase in dataset size. This suggested a ceiling that was not a function of model capacity. UN-P, which shares the identical backbone with $\sim$3.1M additional parameters from cross-attention layers (a 23\% increase), consistently achieved lower validation loss throughout training and converged to a final MSE 13.3\% lower than before. This indicates that the ceiling was an information bottleneck rather than a capacity bottleneck, since the grayscale input simply does not contain enough information to resolve color ambiguity, and text provides the missing signal.

An earlier iteration of the SD tier trained with 15\% empty-prompt dropout and applied classifier-free guidance (CFG) at a scale of 7.5 during inference. This caused the text signal to overwhelm the grayscale conditioning, producing overly vivid and mostly incorrect colors. The final SD models address this by setting the guidance scale to 1.0 for both variants, testing text conditioning through the cross-attention mechanism alone without CFG amplification. While this is a more conservative design that likely underestimates the potential benefit of text conditioning, it provides a cleaner ablation, as any observed difference is attributable to cross-attention conditioning rather than to the multiplicative effect of CFG scaling.

\subsection{Limitations}

Several limitations should be noted. (1) All experiments use a single 7{,}459-image dataset, and results may not generalize to other domains. (2) U-Net predictions ($256^2$) are upscaled to $512^2$ for cross-tier comparison, which penalizes U-Net scores but does not affect within-tier deltas. (3) We rely exclusively on automated metrics, but a user study would strengthen claims about perceptual quality. (4) Both SD variants produce lower colorfulness (0.167--0.168) than the ground-truth target (0.258), likely reflecting VAE decoder or LAB-space properties rather than a text-conditioning failure.

\section{Conclusion}

This paper investigated the extent to which text conditioning improves automatic image colorization when it is the only variable changed. Through controlled ablations across two architecture tiers (a custom U-Net with $\sim$13--16M parameters and a fine-tuned Stable Diffusion 1.5 model with $\sim$859M parameters), we found that text conditioning provides consistent $\sim$5--6\% PSNR gains and meaningful LPIPS reductions at both scales (Table~\ref{tab:cross_tier}). The larger relative improvement in colorfulness in the U-Net tier suggests that text conditioning is most valuable when the model lacks pretrained visual priors, though it still improved the pretrained SD model on every metric.

Several directions could extend these findings. Intermediate CFG scales (2.0--4.0) may yield larger gains than the conservative scale of 1.0 used here, as finding the optimal balance between text influence and grayscale fidelity is an open question. Varying the empty-prompt dropout rate during training could strengthen the unconditional baseline and provide a more nuanced view of text's contribution. Evaluation on larger and more diverse datasets, along with feedback from human judges, would establish whether these improvements transfer to real-world colorization tasks. Finally, a per-category analysis breaking down results by image content (landscapes vs.\ portraits vs.\ objects) could reveal where text conditioning provides the greatest benefit.

\section{Acknowledgement}
\noindent This project was conducted using the ROSIE high-performance computing cluster at the Milwaukee School of Engineering, to which I had access as a freshman. All Stable Diffusion training and inference was performed on an NVIDIA DGX-1 node within ROSIE, a purpose-built deep-learning server delivering up to 960 tensor teraflops. Access to this node was essential because fine-tuning Stable Diffusion 1.5 required approximately 16,GB of GPU memory, which exceeded the 8,GB available on the consumer-grade desktop GPU accessible to the authors. The DGX-1 node reduced each Stable Diffusion training run from what would have been multiple days on consumer hardware to approximately 20 hours (50 epochs at $\sim$25 minutes per epoch), while U-Net training completed in roughly 4--6 hours per variant. Without access to ROSIE, the iterative experimentation across four model variants and multiple architectural revisions necessary for this ablation study would not have been feasible.

\bibliographystyle{plainnat}
\bibliography{references}

\end{document}